# Physical correlations lead to kappa distributions

G. Livadiotis and D.J. McComas

Department of Astrophysical Sciences, Princeton University, Princeton, NJ 08544, USA

**Abstract**

The recently developed concept of "entropic defect" is important for understanding the foundations of thermodynamics in space plasma physics, and more generally, for systems with physical correlations among their particles. Using this concept, this paper derives the basic formulation of the distribution function of velocities (or kinetic energies) in space plasma particle populations. Earlier analyses have shown how the formulation of kappa distributions is interwoven with the presence of correlations among the particles' velocities. This paper shows, for the first time, that the reverse is true: the thermodynamics of particles' physical correlations are consistent only with the existence of kappa distributions.

## 1. Introduction

Space plasmas are collisionless systems with various levels of correlations among their particles, residing in stationary states characterized by a non-Maxwellian behavior, typically described by kappa distributions. Kappa distributions have been employed to describe numerous space plasma populations in the heliosphere, from solar wind and the planetary magnetospheres to the inner heliosheath, and even beyond, to interstellar plasmas. (For more details on the theory of kappa distributions and its connection to nonextensive statistical mechanics and space plasma physics and thermodynamics, see: the books: Livadiotis 2017 and Yoon 2019; the reviews in Livadiotis & McComas 2009; Pierrard & Lazar 2010; Livadiotis & McComas 2013a, Livadiotis 2015a; Tsallis 2019; and, the original applications in: Binsack 1966; Olbert 1968; Vasyliūnas 1968.)

There have been numerous applications of kappa distributions in space plasmas over the past decades. Some examples are: (i) *inner heliosphere*, including solar wind (e.g., Maksimovic et al. 1997; Pierrard et al. 1999; Mann et al. 2002; Marsch 2006; Zouganelis 2008; Štverák et al. 2009; Yoon et al. 2012; Yoon 2014; Pierrard & Pieters 2015; Pavlos et al. 2016; Livadiotis 2018a); solar spectra (e.g., Dzifčáková & Dudík 2013; Dzifčáková et al. 2015; Lörinčík et al. 2020); solar corona (e.g., Vocks et al. 2008; Lee et al. 2013; Cranmer 2014); solar energetic particles (e.g., Xiao et al. 2008; Laming et al. 2013); corotating interaction regions (e.g., Chotoo et al. 2000); solar flares (e.g., Mann et al. 2009; Livadiotis & McComas 2013b; Bian et al. 2014; Jeffrey et al. 2017); solar radio emissions (e.g., Cairns et al. 2004; Li & Cairns 2013; Schmidt & Cairns 2016); (ii) *planetary magnetospheres*, including magnetosheath (e.g., Formisano et al. 1973; Ogasawara et al. 2013); magnetopause (e.g., Ogasawara et al. 2015); magnetotail (e.g., Grabbe 2000); ring current (e.g., Pisarenko et al. 2002), plasma sheet (e.g., Christon 1987; Wang et al. 2003;

Kletzing et al. 2003); magnetospheric substorms (e.g., Hapgood et al. 2011), Aurora (e.g., Ogasawara et al. 2017), magnetospheres of giant planets, such as Jovian (e.g., Collier and Hamilton 1995; Mauk et al. 2004; Nicolaou et al. 2014); Saturnian (e.g., Livi et al. 2014; Carbary et al. 2014; Dialynas et al. 2018); Uranian (e.g., Mauk et al. 1987); magnetospheres of planetary moons, such as Io (e.g., Moncuquet et al. 2002) and Enceladus (e.g., Jurac et al. 2002); cometary magnetospheres (e.g., Broiles et al. 2016; Myllys et al. 2019); (iii) *outer heliosphere and the inner heliosheath* (e.g., Decker et al. 2005; Heerikhuisen et al. 2008; 2015; Zank et al. 2010; Livadiotis et al. 2011; 2012; 2013; 2022; Livadiotis & McComas 2010; 2012; Fuselier et al. 2014; Zirnstein & McComas 2015; Zank, 2015; Livadiotis 2016; Swaczyna et al. 2019); (iv) *beyond the heliosphere*, including HII regions (e.g., Nicholls et al. 2012); planetary nebula (e.g., Nicholls et al. 2013; Zhang et al. 2014; Lin & Zhang 2020; Yao & Zhang 2022); active galactic nuclei (e.g., Humphrey et al. 2019; Morais et al. 2021); galactic jets (e.g., Davelaar et al. 2019); supernova (e.g., Raymond et al. 2010); and, plasmas of cosmological scales (e.g., Hou et al. 2017; Livadiotis & McComas 2021a).

Kappa distributions have been embraced because they provide a strong theoretical framework for the particle distributions commonly observed in space plasmas (and many other systems). Indeed, a breakthrough in the field came with the connection of the kappa distributions with statistical mechanics and thermodynamics (Treumann 1997; Milovanov & Zelenyi 2000; Leubner 2002; Livadiotis & McComas 2009; 2010; Livadiotis 2014; 2015a;b; 2017, Ch.1; 2018b). In particular, among other attributes, kappa distributions: (i) maximize the entropy of nonextensive statistical mechanics under the constraints of canonical ensemble (e.g., Livadiotis & McComas 2009; Livadiotis 2014); (ii) characterize particle systems exchanging heat with each other eventually stabilized into a stationary state defining a non-classical version of thermal equilibrium (generalized thermal equilibrium) (e.g., Abe 2001, Livadiotis 2018a); (iii) constitute the unique description of particle energies consistent with polytropic behavior (e.g., Livadiotis 2018a; 2019a; Nikolaou & Livadiotis 2019); and (iv) lead to a consistent characterization of temperature for systems residing in stationary states out of thermal equilibrium, such as the majority of space plasmas (Livadiotis & McComas 2009; 2010; 2011a; 2013a; 2021b; Livadiotis 2018b).

There are various mechanisms that can generate kappa distributions in plasmas. These particle systems are characterized by local particle correlations and strong collective behavior, leading to a framework of statistical mechanics different from the classical one that is based on the entropy of Boltzmann-Gibbs and the distribution function of Maxwell-Boltzmann. Some examples are: superstatistics (Beck & Cohen 2003; Schwadron et al. 2010; Hanel et al. 2011; Livadiotis et al. 2016; Livadiotis 2019b; Gravanis et al. 2020), effect of shock waves (Zank 2006), weak turbulence (Yoon 2014; 2019), turbulence with a diffusion coefficient inversely proportional to velocity (Bian et al. 2014), effect of pickup ions (Livadiotis & McComas 2011a;b), pump acceleration mechanism (Fisk & Gloeckler 2014), polytropic behavior (Meyer-Vernet et al. 1995; Livadiotis 2018a; 2019a); (see also: Livadiotis 2017, Chapters 4-6,8,10,15,16). Common plasma processes, such as Debye shielding and magnetic coupling, can also play an important role in the generation of kappa distributions in plasmas (Livadiotis et al. 2018).

In general, such physical mechanisms connect the particles through long-range interactions, producing statistical correlations among particles, and leading the system to be stabilized into specific kappa distributions. Previous publications have analytically shown how the formulation of kappa distributions is interwoven with the correlations among the particles' velocities or kinetic energies (Abe 1999; Livadiotis & McComas 2011a; Livadiotis 2015b; Livadiotis et al. 2021). In particular, the factorization property of the exponential function (characterizing the Maxwell-Boltzmann distribution), i.e., $\exp(x+y)=\exp(x)\cdot\exp(y)$, leads to zero correlation among particles. In contrast, the formulation of kappa distributions leads to a nonzero correlation, which is inversely proportional to kappa, the index that parameterizes and labels the kappa distributions.

Nevertheless, in general any non-Maxwellian (i.e., non-exponential) distribution function, and not just the kappa distributions, can lead to finite (nonzero) correlations among particle velocities. Then, we may ask: is the presence of the collective behavior and the particle correlations in space plasmas connected specifically to the kappa distributions, or could it come from any other non-Maxwellian distribution?

Here we address this question, showing the surprising result that, indeed, the presence of physical correlations uniquely implies the existence of kappa distributions. Starting from the existence of correlations between particles in stable and statistically stationary systems, we find that the statistics of the particle kinetics is described only by the specific formulation of kappa distributions. This is addressed by considering the concept of the "entropy defect."

## 2. Entropy Defect
### 2.1. General concept

Recently, we developed the new concept of an entropy defect for thermodynamics, analogous to the well-known mass defect (Figure 1). The mass of the nucleus is less than the sum of the masses of the free constituent of the nucleons (protons and neutrons). The mass defect constitutes the missing mass, that is, the difference between the mass of a composite particle system – the nucleus – and the sum of the masses of its parts, the nucleons. The mass defect is caused by the nuclear binding energy via the Einstein's famous formula, $E = m\,c^2$ (1905), reflecting the energy that was released when the nucleus was formed.

In its thermodynamic analogous, the entropy defect constitutes a missing entropy, that is, the difference between the entropy of a composite particle system – a space plasma – and the sum of the entropies of its parts, the correlated particles. The entropy defect is a measure of the order induced by the existence of particle correlations, reflecting the entropy that was released, when the correlated system was formed. These physical correlations can be caused by various mechanisms that "bind" particles together, such as the coupling of the Debye electrostatic shielding or the embedded magnetic field with the plasma constituents on various temporal and spatial scales.

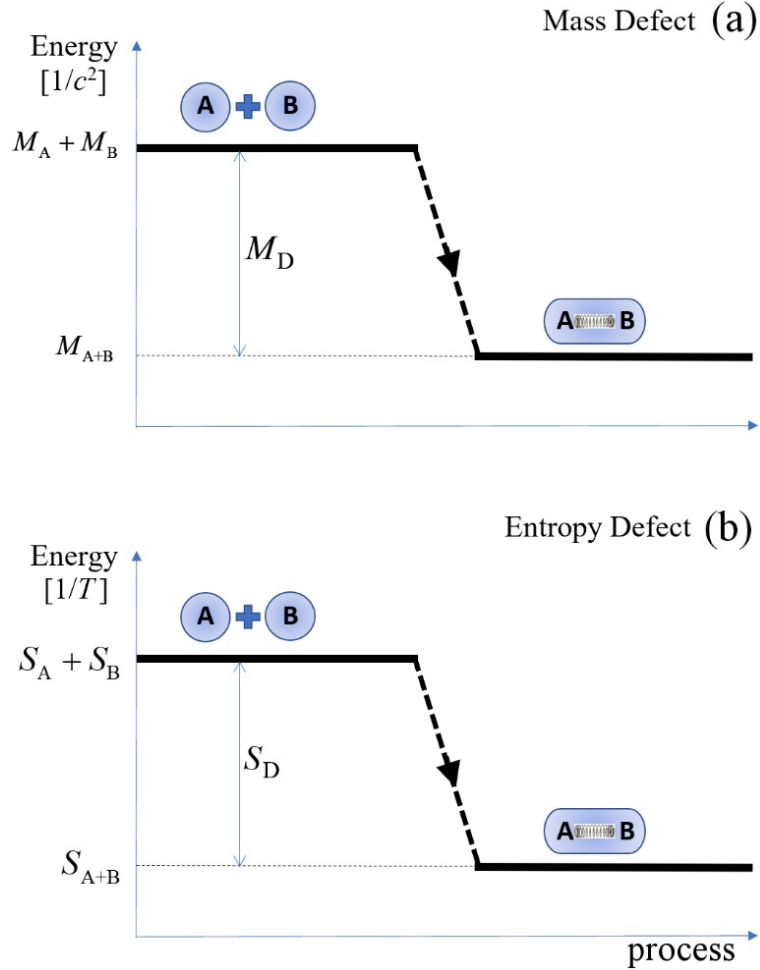

**Figure 1.** Analogy between mass $M_D$ and entropy $S_D$ defects, and their equivalent to an energy level (per $c^2$ and temperature, respectively) change during the composition of a system. (a) The mass defect $M_D$ constitutes the missing mass between the mass of a composite particle system and the sum of the masses of its constituents A and B; it is caused by the release of the binding energy (mass). (b) The entropy defect $S_D$ constitutes the missing entropy between the entropy of a composite particle system and the sum of the entropies of its constituents A and B; it is caused by the order induced by the existence of particle correlations and represents the release of a "binding entropy".

Consider two arbitrary and originally independent systems A and B, interacting through a wall permeable to heat transfer, eventually reaching a stationary thermodynamic state, where no exchange of energy or entropy is observed. The composition of the two subsystems A and B into the combined system A+B is generally completed through an irreversible mixing thermodynamic process, but it can be approached by a progressive series of slow quasi-stationary and quasi-reversible ones. If the subsystems continue to be independent after the composition of the combined system, then, the total entropy is additive (e.g., Lavenda 1978; Schroeder 2000), i.e., $S_{A+B} = S_A + S_B$; however, if correlations develop between the two subsystems during the formation of the combined system, then, the total entropy is nonadditive, i.e., $S_{A+B} \neq S_A + S_B$. The difference between the sum of the subsystems entropies and the total entropy equals a missing amount of entropy, $S_{A+B} = S_A + S_B - \Delta S$; this defines the "entropy defect", namely, the decrease of entropy due to the order generated by the presence of correlations and written as a function of the constituents' entropies (Livadiotis & McComas 2021b),

$$S_{A+B} = S_A + S_B - S_D(S_A, S_B). \tag{1}$$

The concept of entropy defect may be expanded to all the (*N*) particle constituents of a system: the entropy defect equals the amount of entropy, missing between the total entropy of the combined system and the sum of the partition entropies of all the particle constituents of the combined system, $S(N) = S_i(N) - S_D(N)$, where $S_i$ stands for the entropy of the system if there were no correlations among the particles, thus, no entropy defect.

## 2.2. Elementary entropy defect - Thermodynamic definition of Kappa

Let a system of entropy *S* be constructed by combining *N* elementary subsystems (e.g., particles). These are originally independent, but once involved in the combined system, long-range interactions that induce mutual correlations develop between them, adding order to the system, and creating an entropy defect. Adding an elementary independent subsystem of entropy $dS_i$, the change of entropy of the total system would have been $dS = dS_i$, if there were no correlations. However, in the presence of correlations, the entropy defect decreases this entropy change to $dS = dS_i - dS_D$. Applying Eq.(1), we have

$$S + dS = S + dS_i - dS_D(S, dS_i), \tag{2}$$

where the elementary entropy defect $dS_D(S, dS_i)$ has to be proportional to the elementary added entropy, $dS_D \propto dS_i$, and also to the entropy of the combined system, $dS_D \propto S$. Hence,

$$dS_D(S, dS_i) = \frac{1}{\kappa} \cdot S \cdot dS_i. \tag{3}$$

We may consider the general case, where the proportionality parameter depends on the entropy of the system, $\kappa = \kappa(S)$. However, there are multiple lines of direct evidence that the kappa is independent of *S*, so that we do have the proportion $dS_D \propto S$. This is demonstrated through four independent arguments:

*1) Existence of stationary state – generalized thermal equilibrium.*
Previous analyses of the concept of thermal equilibrium revealed that the most generalized entropic rule for mixing originally independent subsystems is governed by Eq.(1) with $S_D(S_A, S_B) \propto S_A \cdot S_B$ (Abe 2001, Livadiotis 2018b;c).

*2) Thermodynamic definition of kappa.*

Equation (3) can be seen as an outcome of the thermodynamic definition of kappa $\kappa$, which, together with the temperature $T$, constitute the two independent intensive parameters characterizing the thermodynamics of (particle) systems residing in stationary states (generalized thermal equilibrium). In particular, the thermodynamic definition of kappa is given by

$$\frac{1}{\kappa} \equiv \frac{1}{S} \cdot \left(\frac{\partial S_\mathrm{D}}{\partial S_\mathrm{i}}\right)_{V,N}, \qquad (4)$$

from where Eq.(3) can be trivially derived. (Further details on the thermodynamic definition of both kappa and temperature can be found in Livadiotis 2018b, Livadiotis & McComas 2021b).

*3) Independence between intensive thermodynamic variables of temperature and kappa.*

The independence of the two parameters can be physically understood by their role in thermodynamic equilibrium. The temperature describes the energy exchange among the constituents of a system, and once the (generalized) thermal equilibrium is achieved, no other energy flow is observed. On the other hand, the kappa describes the partitioning of the entropy of the system into the entropies of its constituents, as this is expressed by Eq.(1), where the entropy or other thermodynamic parameter is clearly not involved.

Nevertheless, several local $\kappa$-$T$ relationships have been observed in space plasmas in the heliosphere and planetary magnetospheres (e.g., Collier & Hamilton 1995; Ogasawara et al. 2013; Broiles et al. 2016; Dialynas et al. 2018; Nicolaou & Livadiotis 2019). Moreover, a characteristic example of such a relationship was observed between the measurements of the kappa and temperature of the proton plasma in the inner heliosheath. Using energetic neutral atom (ENA) observations from the Interstellar Boundary Explorer (IBEX) (McComas et al. 2009a;b) with energies above ~0.7 keV (Funsten et al. 2009), and by connecting the observed ENA flux to the kappa distribution of velocities of the source protons, Livadiotis et al. (2011; 2012; 2013; 2022) produced sky maps of the kappa and temperature of the proton plasma in the inner heliosheath, revealing a trend of positive correlation between these parameters. In any case, the observed $\kappa$-$T$ relationships are not universal but local characteristics of the plasma, interwoven with the specific mechanism that generates kappa distributions in that plasma (Livadiotis et al. 2018).

Finally, the fact that the kappa and temperature are two independent thermodynamic parameters implies that the kappa is not a function of entropy or any other thermodynamic parameter.

*4) Dependence on size in near equilibrium cases.*

The entropy of the elementary subsystem, $dS_\mathrm{ind}$, does not depend on the size of the whole system (e.g., on the number of particles $N$), but does depend on the entropy defect because it characterizes the long-range interactions in the plasma between the elementary subsystem and all the already $N$ combined subsystems. For a sufficiently small entropy defect, the entropy of the combined system is extensive, i.e., $S(N) \propto N$.

On the other hand, for weak correlations, the elementary entropy defect involves the equal interaction of the elementary subsystem with entropy $dS_i$ with each of the $N$ already combined subsystems, so that $dS_D \propto N$. This remark supports the fact that the proportion $dS_D \propto S$ is realistic for physical systems.

## 3. Generalized Sackur-Tetrode equation, derived from the concept of entropy defect

Here we derive the generalized Sackur-Tetrode equation, that is, the entropy expressed as a function of thermal variables, which is applicable for non-Maxwellian distributions in space plasmas. The classical Sackur–Tetrode equation is the similar expression of entropy as a function of thermal variables, but under the Boltzmann-Gibbs statistical mechanics and Maxwell-Boltzmann distribution (Sackur 1911; Tetrode 1912).

We derive the entropy $S$ as a function of thermal variables in the following three steps. First, we find the entropy $S$ as a function of the entropy in the absence of correlations, $S_i$, i.e., $S=F(S_i)$; second, we express the entropy in the absence of correlations $S_i$ as a function of thermal variables, i.e., $S_i = G(T,...)$. Finally, the combination of the two functions leads to the desired expression of $S=(F \circ G)(T,...)$.

*(a) Derivation of $S=F(S_i)$.*

The combination of Eqs.(2,3) gives $dS = (1 - \frac{1}{\kappa} \cdot S) \cdot dS_i$, or

$$(1 - \tfrac{1}{\kappa} \cdot S)^{-1} \cdot dS = dS_i, \tag{5a}$$

which after the integration of the combined system's entropy from 0 to $S$, and the summation of the elementary subsystems (combined system's entropy in the absence of correlations) from 0 to $S_i$, we find $\ln(1 - \tfrac{1}{\kappa} \cdot S)^{-\kappa} = S_i$, or

$$\exp_\kappa(S) \equiv (1 - \tfrac{1}{\kappa} \cdot S)^{-\kappa} = \exp(S_i), \tag{5b}$$

or

$$S_i = \ln[\exp_\kappa(S)] \equiv \ln[(1 - \tfrac{1}{\kappa} \cdot S)^{-\kappa}], \tag{6a}$$

$$S = \ln_\kappa[\exp(S_i)] \equiv \kappa \cdot [1 - \exp(-\tfrac{1}{\kappa} \cdot S_i)], \tag{6b}$$

where we have used the $\kappa$-deformed exponential function, and its inverse, the deformed logarithm function (e.g., see: Nivanen et al. 2003; Borges 2004; Suyari 2006),

$$\exp_\kappa(x) \equiv (1 - \tfrac{1}{\kappa} x)^{-\kappa}, \quad \ln_\kappa(x) \equiv \kappa(1 - x^{-\frac{1}{\kappa}}), \tag{7a}$$

with

$$\exp_\kappa[\ln_\kappa(x)] = \ln_\kappa[\exp_\kappa(x)] = x. \tag{7b}$$

*(b) Derivation of $S_i = G(T,...)$.*

The entropy in the absence of correlations, $S_i$, is simply described by the classical Sackur-Tetrode equation, given as a function of temperature,

$$S_i = \tfrac{1}{2} d \cdot N \cdot \ln(T/T_0), \qquad (8)$$

where the thermal constant $T_0$ constitutes the minimum temperature for the entropy to be positive.

*(c) Derivation of $S=(F \circ G)(T,...)$.*

Combining the expression of the extensive entropy in terms of the temperature, $S_i(T)$, with the expression of the nonextensive $S$ in terms of extensive $S_i$ entropies, we end up with the expression of the nonextensive entropy $S$ in terms of the temperature, $S(T)$,

$$S = \kappa \cdot \left(1 - e^{-\frac{1}{\kappa} \cdot S_i}\right) = \kappa \cdot \left[1 - (T/T_0)^{-\frac{1}{\kappa} \cdot \frac{1}{2} d \cdot N}\right], \qquad (9a)$$

or, using the convenient formalism of the $\kappa$-deformed functions

$$S \equiv \ln_\kappa \left[\exp(S_i)\right] = \ln_\kappa \left[(T/T_0)^{\frac{1}{2} d \cdot N}\right]. \qquad (9b)$$

The involved thermal constant is given by $k_B T_0 = C \cdot \hbar_c^2 (m_e m_i)^{-\frac{1}{2}} \cdot \lambda_c^{-2} \cdot g_{\kappa,N}$, where $C=(9\pi/2)^{1/3}/e \approx 0.89$ (Livadiotis & McComas 2013b; Livadiotis 2014; 2017, Chapters 2 and 5); $\lambda_c$ is the smallest correlation length, which is interpreted by the interparticle distance $b \sim n^{-1/d}$ for collisional particle systems (absence of correlations), or by the Debye length $\lambda_D$ for collisionless particle systems (presence of local correlations) (Bryant 1996; Rubab & Murtaza 2006; Gougam & Tribeche 2011; Livadiotis & McComas 2014a; Livadiotis 2019c; Saberian 2019); the factor $g_{\kappa,N}$ depends on the kappa index $\kappa$ and the number of correlated particles; for large number of particles, it becomes $g_{\kappa,N} \approx 1$; finally, the phase-space parcel $\hbar_c$ is typically given by the Planck's constant, but it was shown to represent a different and larger constant in space plasmas, where Debye shielding limits the distance of correlations, $\hbar_* = (1.19 \pm 0.05) \times 10^{-22}\,\text{J}\cdot\text{s}$ (Livadiotis & McComas 2013b; 2014b; Scholkmann 2013; Livadiotis & Desai 2016; Livadiotis 2019d; Livadiotis et al. 2020); (namely, $\hbar_c = \hbar$, when no correlations exist among particles, and $\hbar_c = \hbar_*$, when correlations exist among particles beyond their nearest neighbors, as in the general case of space plasmas).

In addition, we may provide the entropy as a function of the inverse temperature $\beta \equiv (k_B T)^{-1}$ (and $\beta_0 \equiv (k_B T_0)^{-1}$), as this will be useful in the next section,

$$S = \kappa \cdot \left[1 - (\beta/\beta_0)^{\frac{1}{\kappa} \cdot \frac{1}{2} d \cdot N}\right] = \ln_\kappa \left[(\beta/\beta_0)^{-\frac{1}{2} d \cdot N}\right]. \qquad (10)$$

## 4. The general Sackur-Tetrode equation of entropy leads to the kappa distributions of velocities

The Sackur-Tetrode equation expresses the entropy as a function of the thermal variables, $S(\beta/\beta_0)$. Typically, this is derived from the entropic statistical equation, $S(p)$, that is, the entropy expression in terms of the probability, after substituting the probability distributed with energy, which is expressed in terms of thermal observables, $p(\varepsilon, \beta/\beta_0)$, and then, integrating over energy, $\int S[p(\varepsilon, \beta/\beta_0)]g(\varepsilon)d\varepsilon$, where $g(\varepsilon) \propto \varepsilon^{\frac{1}{2}d-1}$ is the density of states in a *d*-dimensional velocity space. On the other hand, in this paper, we have derived the Sackur-Tetrode equation without passing through the entropic statistical equation, but instead, starting from the concept of entropy defect. From this, it is straightforward to reverse these steps to determine the entropy in terms of the probability distribution function once it is known in terms of the thermal observables – the entropic statistical equation.

One may think that there are two unknowns in the reverse procedure, i.e., the statistical equation of entropy and the probability distribution function. However, these do not constitute independent unknowns, since, the distribution function can be determined from the statistical equation of entropy and its maximization under the constraints of the canonical ensemble. Here, we derive the statistical equation of entropy (i.e., the entropy as a function of probabilities), given the Sackur-Tetrode equation of entropy (i.e., the entropy as a function of the thermal variables), derived in the previous section using the concept of the entropy defect.

Therefore, starting from the entropy defect we derive the statistical equation of entropy (Figure 2). Furthermore, since the entropy is connected to the distribution function (through the maximization of entropy under the constraints of the canonical ensemble), the entropy defect leads to the distribution function of energy that characterizes the particle system.

We come to these derivations, through the following steps: a) express the statistical equation of entropy via an unknown function of probabilities, $f(p)$; (b) connect the energy distribution function, $p(\varepsilon)$, through the same unknown function $f$; (c) connect the derived Sackur-Tetrode equation with the statistical equation of entropy that involves the unknown function $f$; and (d) determine the specific functional form of $f$, which is consistent to the entropy defect.

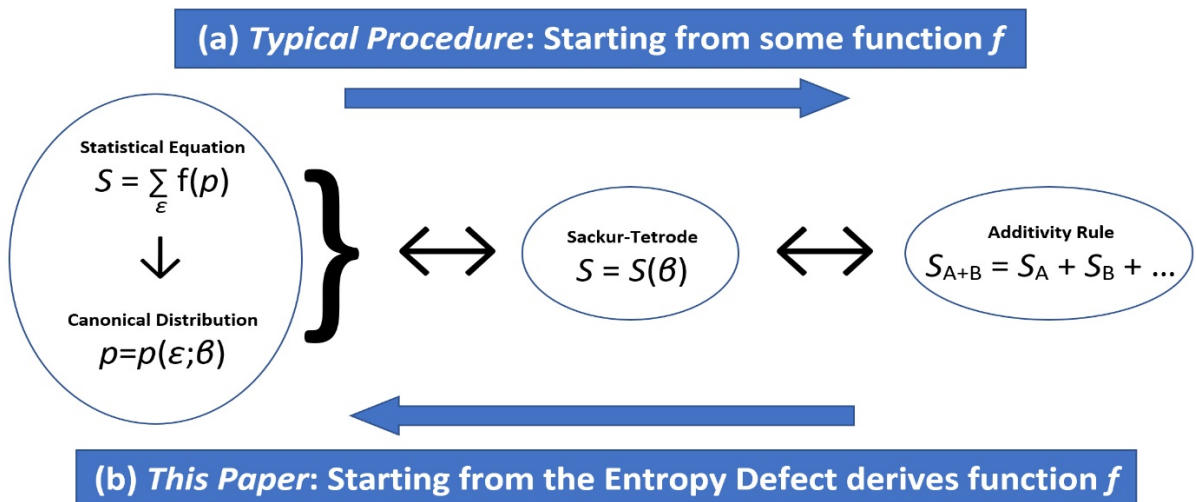

**Figure 2.** The scheme demonstrates the method used in this paper to derived the entropic function.

*(a) General expression of the entropic statistical function*

In space plasmas, the particle energy spectrum and its distribution function $p(\varepsilon)$ are considered continuous. Nevertheless, for sake of simplicity, we switch to a discrete formalism, where the energy spectrum $\{\varepsilon_k\}_{k=1}^{W}$ (with $k=1, \ldots, W$) is associated with a discrete probability distribution $\{p_k\}_{k=1}^{W}$. The general entropic form is still a general function of the probabilities, $S = S(\{p_k\}_{k=1}^{W})$. However, under the following reasonable physical requirements of the entropy: (1) it can be maximized, and (2) it does not depend on the states themselves but only on their assigned probabilities, and thus it is symmetric to any permutation of each components, e.g., $S = S(\ldots, p_k, \ldots, p_\ell, \ldots) = S(\ldots, p_\ell, \ldots, p_k, \ldots)$, then, we can write

$$S = \sum_{k=1}^{W} f(p_k), \tag{11}$$

(where we set the Boltzmann constant $k_B$ to 1). For example, in the cases of Gibbs (1902) and Tsallis (1988) entropies, the function $f$ is respectively given by:

$$f(x) = -x\ln(x) \quad \text{and} \quad f(x) = (x - x^q)/(q-1). \tag{12}$$

*(b) Connection between the entropy and distribution functions*

Next, we connect the entropic statistical function $f(p)$ with the canonical probability distribution function $p(\varepsilon)$, so that knowing either of the two functions, the other one can be derived; (a similar treatment can be found in Livadiotis 2018d).

The maximization of entropy under the constraints of canonical ensemble, i.e., $1 = \sum_{k=1}^{W} p_k$ (normalization) and $U = \sum_{k=1}^{W} p_k \varepsilon_k$ (internal energy), involves maximizing the functional $G(\{p_k\}_{k=1}^{W}) = \sum_{k=1}^{W} f(p_k) + \lambda_1 \sum_{k=1}^{W} p_k + \lambda_2 \sum_{k=1}^{W} p_k \varepsilon_k$. Hence, setting $\partial G(\{p_k\}_{k=1}^{W})/\partial p_i = 0$, we obtain

$$f'(p_i) + \lambda_1 + \lambda_2 \varepsilon_i = 0, \quad \text{or} \quad p_i(\varepsilon_i) = f'^{-1}(-\lambda_1 - \lambda_2 \varepsilon_i), \tag{13}$$

where the coefficients $\lambda_1$ and $\lambda_2$ are the Lagrangian multipliers. The (negative) second Lagrangian multiplier, $-\lambda_2$, in the canonical ensemble represents the inverse temperature $\beta \equiv (k_B T)^{-1}$, or, $-\lambda_2 = k \cdot \beta$, where $k$ is a proportionality parameter independent of temperature. In the continuous description we have

$$p(\varepsilon) \propto f'^{-1}(-\lambda_1 - \lambda_2 \varepsilon) = f'^{-1}(k \cdot \beta \varepsilon - \lambda_1), \tag{14}$$

with normalization $1 = \int_0^\infty p(\varepsilon) g(\varepsilon) d\varepsilon$, where the density of states is typically given by $g(\varepsilon) = g_1 \varepsilon^{\frac{1}{2}d-1}$, where $g_1$ includes all the involved constants and $d$ is the dimensionality of the velocity space. Therefore,

$$p(\varepsilon) = A^{-1} k^{\frac{1}{2}d} \cdot \beta^{\frac{1}{2}d} \cdot f'^{-1}(k \cdot \beta \varepsilon - \lambda_1), \tag{15}$$

where the normalization constant $A$ is independent of temperature and can be determined from

$$1 = A^{-1} g_1 k^{\frac{1}{2}d} \cdot \beta^{\frac{1}{2}d} \cdot \int_0^\infty f'^{-1}(k \cdot \beta \varepsilon - \lambda_1) \varepsilon^{\frac{1}{2}d-1} d\varepsilon. \tag{16}$$

*(c) Connection between the Sackur-Tetrode and the statistical equations of entropy*

The entropic sum in Eq.(11) can be written in the continuous description as

$$S = \int_0^\infty f[p(\varepsilon)]g(\varepsilon)d\varepsilon = g_1 \cdot \int_0^\infty f[A^{-1}k^{\frac{1}{2}d} \cdot \beta^{\frac{1}{2}d} \cdot f'^{-1}(k \cdot \beta\varepsilon - \lambda_1)]\varepsilon^{\frac{1}{2}d-1}d\varepsilon. \tag{17}$$

Setting $x \equiv k\beta\varepsilon$, we obtain

$$S = \beta^{-\frac{1}{2}d} \cdot k^{-\frac{1}{2}d} g_1 \cdot \int_0^\infty f\left[A^{-1}k^{\frac{1}{2}d} \cdot \beta^{\frac{1}{2}d} \cdot f'^{-1}(x - \lambda_1)\right]x^{\frac{1}{2}d-1}dx. \tag{18}$$

Then, we substitute in Eq.(18) the Sackur-Tetrode entropy equation, $S=S(\beta)$, as derived from the entropy defect and shown in Eq.(10) (for $N=1$), i.e., $S = \kappa \cdot \left[1-(\beta/\beta_0)^{\frac{1}{\kappa}\cdot\frac{1}{2}d}\right]$. Namely,

$$\kappa \cdot \left\{1 - \left[(\beta/\beta_0)^{\frac{1}{2}d}\right]^{\frac{1}{\kappa}}\right\} = (\beta/\beta_0)^{-\frac{1}{2}d} \cdot g_1 \cdot \int_0^\infty f[A^{-1} \cdot (\beta/\beta_0)^{\frac{1}{2}d} \cdot f'^{-1}(x-\lambda_1)]x^{\frac{1}{2}d-1}dx, \tag{19}$$

where we have redefined the constants $A^{-1}k^{\frac{1}{2}d}\beta_0^{\frac{1}{2}d} \to A^{-1}$ and $g_1 k^{-\frac{1}{2}d}\beta_0^{-\frac{1}{2}d} \to g_1$.

*(d) Derivation of the entropy function consistent to the entropy defect*

In this final step, we compare the two Sackur-Tetrode equations, (i) the one derived using the entropy defect, shown in Eq.(10), and (ii) the one expressed in terms of an unknown entropy function $f$, as in Eq.(18), in order to determine the function $f$, which is involved in the entropy $S$, as shown in Eq.(11), and in the distribution function $p$, as shown in Eq.(15). Starting from Eq.(19), we set $y \equiv A^{-1} \cdot (\beta/\beta_0)^{\frac{1}{2}d}$, i.e.,

$$\kappa A \cdot y - \kappa A^{1+\frac{1}{\kappa}} \cdot y^{1+\frac{1}{\kappa}} = g_1 \cdot \int_0^\infty f\left[y \cdot f'^{-1}(x-\lambda_1)\right] \cdot x^{\frac{1}{2}d-1}dx. \tag{20}$$

Differentiating in terms of $y$, and multiplying by $y$, we obtain

$$\kappa A \cdot y - (\kappa+1)A^{1+\frac{1}{\kappa}} \cdot y^{1+\frac{1}{\kappa}} = g_1 \cdot \int_0^\infty f'\left[y \cdot f'^{-1}(x-\lambda_1)\right] \cdot y \cdot f'^{-1}(x-\lambda_1) \cdot x^{\frac{1}{2}d-1}dx. \tag{21}$$

Combining Eqs.(20,21), we have

$$A^{1+\frac{1}{\kappa}} \cdot y^{1+\frac{1}{\kappa}} = g_1 \cdot \int_0^\infty \left\{f\left[y \cdot f'^{-1}(x-\lambda_1)\right] - f'\left[y \cdot f'^{-1}(x-\lambda_1)\right] \cdot y \cdot f'^{-1}(x-\lambda_1)\right\} \cdot x^{\frac{1}{2}d-1}dx. \tag{22}$$

The left-hand side can be replaced by its equal, $(Ac^{-1})^{1+\frac{1}{\kappa}} \cdot g_1 \cdot \int_0^\infty \left[y \cdot f'^{-1}(x-\lambda_1)\right]^{1+\frac{1}{\kappa}} \cdot x^{\frac{1}{2}d-1}dx$, where we set

$$c \equiv \left[g_1 \cdot \int_0^\infty f'^{-1}(x-\lambda_1)^P \cdot x^{\frac{1}{2}d-1}dx\right]^{1/P} \text{ with } P = 1+\tfrac{1}{\kappa}. \tag{23}$$

Therefore, we arrive at

$$0 = \int_0^\infty G[Z(x,y)] \cdot x^{\frac{1}{2}d-1}dx, \text{ with } Z(x,y) \equiv y \cdot f'^{-1}(x-\lambda_1), \text{ and} \tag{24}$$

$$G(x) \equiv f(x) - x \cdot f'(x) - (Ac^{-1})^{1+\frac{1}{\kappa}} \cdot x^{1+\frac{1}{\kappa}}. \tag{25}$$

The only way for the integral to be zero for any arbitrary value of $y$, or of $Z(\underline{x},y)$, is the integrated function $G$ to be zero. Hence, $G(x) = 0$, or

$$f(x) - x \cdot f'(x) - (Ac^{-1})^{1+\frac{1}{\kappa}} \cdot x^{1+\frac{1}{\kappa}} = 0 \,. \tag{26}$$

This is a standard first order differential equation with solution

$$f(x) = \left[ f(1) + \kappa (Ac^{-1})^{1+\frac{1}{\kappa}} \right] \cdot x - \kappa (Ac^{-1})^{1+\frac{1}{\kappa}} \cdot x^{1+\frac{1}{\kappa}} \,. \tag{27}$$

The entropic function must obey $f(0) = 0$, $f(1) = 0$, because the entropy is zero when there is only one possibility (Livadiotis 2018d). Then, Eq.(27) becomes

$$f(x) = (Ac^{-1})^{1+\frac{1}{\kappa}} \cdot \kappa \cdot (x - x^{1+\frac{1}{\kappa}}) \,. \tag{28}$$

In Appendix A, we show that the proportionality constant is $Ac^{-1} = 1$; hence, we arrive at

$$f(x) = \kappa \cdot (x - x^{1+\frac{1}{\kappa}}) \,. \tag{29}$$

This may be rewritten using the index $q$, defined by

$$q \equiv 1 + 1/\kappa \iff \kappa \equiv 1/(q-1) \,, \tag{30}$$

i.e, we find

$$f(x) = x \cdot (1 - x^{q-1})/(q-1) = (x - x^q)/(q-1) \,. \tag{31}$$

Therefore, the most general entropic function, as shown in Eq.(11) is given by

$$S = \sum_{k=1}^{W} \frac{p_k - p_k^q}{q-1} = \frac{1 - \sum_{k=1}^{W} p_k^q}{q-1} \,, \tag{32}$$

where we used the probability normalization $\sum_{k=1}^{W} p_k = 1$.

We end up with the standard nonextensive formalism, where the entropic function can be equivalently expressed in terms of $q$ or $\kappa$,

$$S = \frac{1 - \sum_{k=1}^{W} p_k^q}{q-1} \quad \text{or} \quad S = \kappa \cdot \left( 1 - \sum_{k=1}^{W} p_k^{1+1/\kappa} \right) , \tag{33a}$$

that corresponds to the Boltzmann-Gibbs entropic function (Gibbs 1902), for $\kappa \to \infty$,

$$S = -\sum_{k=1}^{W} p_k \ln p_k \,. \tag{33b}$$

Furthermore, this is maximized under the constraints of canonical ensemble, leading to the kappa distribution function (e.g., Livadiotis & McComas 2009; Livadiotis 2014).

## 5. Conclusions

This paper used the recently developed concept of entropy defect to resolve a key-challenge in the study of the thermodynamics of space plasmas, and in general, of particle systems out of the classical thermal equilibrium. We addressed the question of the basic formulation of the distribution function that characterizes the velocities (or kinetic energies) of space plasma particle populations. Numerous observations have shown the non-Maxwellian behavior of these distributions, with kappa distributions broadly characterizing the observed non-Maxwellian distribution functions. Indeed, kappa distributions have become increasingly found across the physics of space plasma processes, describing particles in the heliosphere from the solar wind and planetary magnetospheres to the inner heliosheath and beyond, and even out into the interstellar plasmas.

Earlier analyses have shown how the formulation of kappa distributions is interwoven with the correlations among the particles' velocities or kinetic energies. This paper used the entropy defect to show, for the first time, the reverse result: the correlations among particles' velocities can be consistent only with the formulation of kappa distributions (Figure 3). This was demonstrated in three steps: (i) connected the particle correlations with the concept of entropy defect, namely, the decrease of entropy due to the order generated by the presence of correlations, (ii) derived the entropy as a function of kappa and temperature, a generalized version of the Sackur-Tetrode equation, and (iii) showed that entropic function derived in (ii) is uniquely expressed by the formalism of kappa distributions of velocities (or kinetic energies).

Therefore, thermodynamics of space plasmas is governed uniquely by kappa distributions. Thermodynamics determines how a particle system behaves when it resides at thermal equilibrium —the concept that any flow of heat is in balance. In that case, the system of particles resides in the final stationary state, assigned only by a temperature, where particle velocities (or kinetic energies) are stabilized into a Maxwell (Boltzmann) distribution. Similarly, space thermodynamics determines how plasma particles behave when their plasma population resides at the concept of generalized thermal equilibrium, the set of stationary states assigned by the independent variables of both temperature and kappa, where particle velocities (or kinetic energies) are stabilized into a kappa distribution.

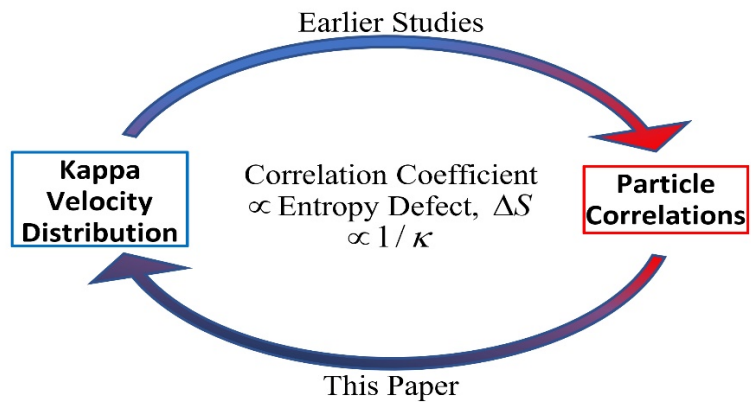

**Figure 3.** The scheme demonstrates the contribution of this paper in understanding the role and importance of kappa distribution function in space plasma physics. Earlier studies have shown the correlations in particle velocities induced by the specific formulation of kappa distributions (Abe 1999; Livadiotis & McComas 2011a; Livadiotis 2015b; Livadiotis et al. 2021). Here, we showed for the first time the reverse result: the presence of correlations among particles residing in stationary states cannot be the outcome of any non-Maxwellian distribution but it is only connected to kappa distributions.

This paper used the concept of entropy defect in space thermodynamics to prove, for the first time, that the kappa distributions uniquely describe the particle velocities in space plasmas, as well as in other particle systems with correlations, residing in stationary states out of the classical thermal equilibrium. With the results provided in this study, it is straightforward that the kappa function constitutes the only right distribution that must be used for describing the particle velocities in space plasma throughout the heliosphere and inner heliosheath, and thus, for analysing the datasets from IBEX and IMAP missions.

This work was funded by the IBEX mission as part of NASA's Explorer Program (80NSSC18K0237) and IMAP mission as a part of NASA's Solar Terrestrial Probes (STP) Program (80GSFC19C0027).

**References**

Abe, S. 1999, PhysA, 269, 403
Abe, S. 2001, PRE, 63, 061105
Beck, C., & Cohen, E.G.D 2003, PhysA, 322, 267
Bian, N., Emslie, G.A., Stackhouse, D.J., & Kontar E.P. 2014, ApJ, 796, 142
Binsack, J.H. 1966, Plasma Studies with the IMP-2 Satellite. Ph.D. Thesis, MIT, Cambridge, USA
Borges, E.P. 2004, Physica A, 340, 95
Broiles, T.W., Livadiotis, G., Burch, J.L., et al. 2016, JGR, 121, 7407
Bryant, D. A. 1996, J. Plasma Phys, 56, 87
Cairns, I. H., Mitchell, J.J., Knock, S.A., Robinson, P.A. 2004, ASR, 34, 88
Carbary, J. F., Kane, M., Mauk, B. H., & Krimigis, S. M. 2014, JGR, 119, 8426
Chotoo, K., Schwadron, N., Mason, G., et al. 2000, JGR, 105, 23107
Christon, S. P. 1987, Icarus, 71, 448
Collier, M.R., & Hamilton, D.C. 1995, GRL, 22, 303
Cranmer, S. R. 2014, ApJL, 791, L31
Davelaar, J., Olivares, H., Porth, O., et al. 2019, A&A, 632, A2
Decker, R. B., Krimigis, S. M., Roelof, E. C., et al. 2005, Science, 309, 2020
Dialynas, K., Roussos, E., Regoli, L.; et al. 2018, JGRA, 123, 8066
Dzifčáková, E., & Dudík, J. 2013, ApJS, 206, 6
Dzifčáková, E., Dudík, J., Kotrč, P., Fárník, F., & Zemanová, A. 2015, ApJS, 217, 14
Einstein, A. 1905, Annalen der Physik, 18, 639
Fisk, L.A., & Gloeckler, G. 2014 JGRA, 119, 8733
Formisano, V., Moreno, G., Palmiotto, F., & Hedgecock, P. C. 1973, JGR, 78, 3714
Fuselier, S. A., Allegrini, F., Bzowski, M., et al. 2014, ApJ, 784, 89
Gibbs, J. W. 1902 Elementary Principles in Statistical Mechanics (New York: Scribner's sons)
Gougam, L. A., & Tribeche, M. 2011, Phys. Plasmas, 18, 062102
Grabbe, C. 2000, PRL, 84, 3614
Gravanis, E., Akylas, E., & Livadiotis, G. 2020, EPL, 130, 30005
Hanel, R., Thurner, S., & Gell-Mann, M. 2011, PNAS, 108, 6390
Hapgood, M., Perry, C., Davies, J., & Denton, M. 2011, Planet. Space Sci. 59, 618
Heerikhuisen, J., Pogorelov, N. V., Florinski, V., Zank, G. P., & le Roux, J. A. 2008, ApJ, 682, 679
Heerikhuisen, J., Zirnstein, E., & Pogorelov, N. 2015, JGR, 120, 1516
Hou, S. Q., He, J. J., Parikh, A., Kahl, D., Bertulani, C. A., Kajino, T., Mathews, G. J., & Zhao, G. 2017, ApJ, 834, 165
Humphrey, A., Villar-Martín, M., Binette, L., Raj, R. 2019, A&A, 621, A10, 14pp
Jeffrey, N. L. S., Fletcher, L., & Labrosse, N. 2017, ApJ, 836, 35

Jurac, S., McGrath, M. A., Johnson, R. E., Richardson, J. D., Vasyliunas, V. M., & Eviatar, A. 2002, GRL, 29, 2172
Kletzing, C. A., Scudder, J. D., Dors, E. E., & Curto, C. 2003, JGR, 108, 1360
Laming, J. M., Moses, J. D., Ko, Y.-K., Ng, C. K., Rakowski, C. E., & Tylka, A. J. 2013, ApJ, 770, 73
Lavenda, B. H. 1978, Thermodynamics of Irreversible Processes - Macmillan Press, London.
Lee, E., Williams, D. R., & Lapenta, G. 2013, arXiv:1305.2939v
Leubner, M.P. 2002, Ap&SS, 282, 573
Li, B. & Cairns, Iver H. 2013
Lin, B.Z., & Zhang, Y. 2020, ApJ, 899, 33
Livadiotis, G., & Desai, M. I. 2016, ApJ, 829, 88
Livadiotis, G., & McComas, D. J. 2009, JGRA, 114, 11105
Livadiotis, G., & McComas, D. J. 2010, ApJ, 714, 971
Livadiotis, G., & McComas, D. J. 2011a, ApJ, 741, 88
Livadiotis, G., & McComas, D. J. 2011b, ApJ, 738, 64
Livadiotis, G., & McComas, D. J. 2012, ApJ, 749, 11
Livadiotis, G., & McComas, D. J. 2013a, SSRv, 75, 183
Livadiotis, G., & McComas, D. J. 2013b, Entrp, 15, 1118
Livadiotis, G., & McComas, D. J. 2014a, JPlPh, 80, 341
Livadiotis, G., & McComas, D. J. 2014b, JGRA, 119, 3247
Livadiotis, G., & McComas, D. J. 2021a, EPL, 135, 49001
Livadiotis, G., & McComas, D. J. 2021b, Entrp, 23, 1683
Livadiotis, G. 2014, Entrp, 16, 4290
Livadiotis, G. 2015a, JGRA, 120, 1607
Livadiotis, G. 2015b, Entrp, 17, 2062
Livadiotis, G. 2016, ApJS, 223, 13
Livadiotis, G. 2017, Kappa Distribution: Theory Applications in Plasmas (1st ed.; Netherlands: Elsevier)
Livadiotis, G. 2018a, JGRA, 123, 1050
Livadiotis, G. 2018b, EPL, 122, 50001.
Livadiotis, G. 2018c, JPCS, 1100, 012017
Livadiotis, G. 2018d, Nonlin Processes Geophys, 25, 77
Livadiotis, G. 2019a, ApJ, 874, 10
Livadiotis, G. 2019b, ApJ, 886, 3
Livadiotis, G. 2019c, Phys. Plasmas 26, 050701
Livadiotis, G. 2019d, ApJ, 887, 117
Livadiotis, G., McComas, D. J, Dayeh, M. A., Funsten, H. O., & Schwadron, N. A. 2011, ApJ, 734, 1
Livadiotis, G., McComas, D. J., Randol, B., et al. 2012, ApJ, 751, 64
Livadiotis, G., McComas, D. J., Schwadron, N. A., Funsten, H. O., & Fuselier, S. A. 2013, ApJ, 762, 134
Livadiotis, G., Assas, L. Dennis, B., Elaydi, S., & Kwessi, E. 2016, NRM, 29, 130
Livadiotis, G., Desai, M.I., & Wilson III, L.B. 2018, ApJ, 853, 142
Livadiotis, G., Dayeh, M. A., Zank, G. P. 2020, 905 (2), 137
Livadiotis, G., Nicolaou, G., & Allegrini, F. 2021, ApJ, 253, 16
Livadiotis, G., McComas, D. J., Funsten, H. O., Schwadron, N.A., Szalay, J. R., & Zirnstein, E. 2022, ApJSS, in Press
Livi, R., Goldstein, J., Burch, J. L., Crary, F., Rymer, A. M., Mitchell, D. G., & Persoon, A. M. 2014, JGRA, 119, 3683
Lörinčík, J., Dudík, J., del Zanna, G., Dzifčáková, E., Mason, H. E. 2020, ApJ, 893, 34
Maksimovic, M., Pierrard, V., & Lemaire, J. 1997, A&A, 324, 725
Mann, G., Classen, H. T., Keppler, E., & Roelof, E. C. 2002, A&A, 391, 749
Mann, G., Warmuth, A., & Aurass, H. 2009, A&A, 494, 669.
Marsch, E. 2006, Living Rev. Solar Phys., 3, 1

Mauk, B. H., Krimigis, S. M., Keath, E. P., Cheng, A. F., Armstrong, T. P., Lanzerotti, L. J., Gloeckler, G., Hamilton, D. C. 1987, JGR, 92, 15283
Mauk, B. H., Mitchell, D. G., McEntire, R. W., Paranicas, C. P., Roelof, E. C., Williams, D. J., Krimigis, S. M., & Lagg, A. 2004, JGRA, 109, A09S12
McComas, D.J., Allegrini, F., et al. 2009a, SpScR, 146, 11
McComas, D. J., Allegrini, F., Bochsler, P., et al. 2009b, Science, 326, 959
Meyer-Vernet, N., Moncuquot, M., & Hoang, S. 1995, Icarus, 116, 202
Milovanov, A.V., & Zelenyi, L.M. 2000, Nonlinear Process. Geophys., 7, 211
Moncuquet, M., Bagenal, F., & Meyer-Vernet, N. 2002, JGRA, 108, 1260
Morais, S. G., Humphrey, A., Villar Martín, M., Binette, L., & Silva, M. 2021, MNRAS, 506, 1389
Myllys, M., Henri, P., & Galand, M. 2019, A & A, 630, A42
Nicholls, D. C., Dopita, M. A., Sutherland, R. S. 2012, ApJ, 752, 148
Nicholls, D. C., Dopita, M. A., Sutherland, R. S., Kewley, L. J., Palay, E. 2013, ApJS, 207, 21
Nicolaou, G., & Livadiotis, G. 2019, 884, 52
Nicolaou, G., Livadiotis, G., & Moussas, X. 2014, SoPh, 289, 1371
Nivanen, L., Le Mehaute, A., & Wang, Q.A. 2003, Rep. Math. Phys., 52, 437.
Ogasawara, K., Angelopoulos, V., Dayeh, M.A., Fuselier, S.A., Livadiotis, G., McComas, D.J., & McFadden, J.P. 2013, JGRA, 118, 3126
Ogasawara, K., Dayeh, M.A., Funsten, H.O., Fuselier, S.A., Livadiotis, G., & McComas, D.J. 2015, JGR, 120, 964.
Ogasawara, K., Livadiotis, G., Grubbs, G.A., Jahn, J.-M., Michell, R., Samara, M., Sharber, J. R., & Winningham, J. D., 2017, GRL, 44, 3475
Olbert, S., 1968, Summary of experimental results from M.I.T. detector on IMP-1, In Physics of the Magnetosphere; Carovillano,
Pavlos, G.P., Malandraki, O.E., Pavlos, E.G., Iliopoulos, A.C., Karakatsanis, L.P. 2016, PhysA, 464, 149
Pierrard, V., & Lazar, M. 2010, SoPh, 267, 153
Pierrard, V., & Pieters, M. 2015, JGRA, 119, 9441
Pierrard, V., Maksimovic, M., & Lemaire, J. 1999, JGR, 104, 17021
Pisarenko, N. F., Budnik, E. Yu., Yermolaev, Yu. I., Kirpichev, I. P., Lutsenko, V. N., Morozova, E. I., Antonova, E. E. 2002, J. Atm. Solar-Terr. Phys. 64, 573
Raymond, J. C., Winkler, P. F., Blair, W. P., Lee, J.-J., & Park, S. 2010, ApJ, 712, 901
Rubab, N., & Murtaza, G. 2006, Phys. Scr., 74, 145.
Saberian, E. 2019, ApJ 887 121
Sackur, O. 1911, Ann. Phys., 36, 958
Schmidt, J. M. & Cairns, I. H. 2016, GRL, 43, 50
Scholkmann, F. 2013, Prog. Phys., 4, 85
Schroeder, D. An Introduction to Thermal Physics; Addison Wesley Longman: USA, 2000, p.20–21
Schwadron, N.A., Dayeh, M., Desai, M., Fahr, H., Jokipii, J.R., & Lee, M.A. 2010, ApJ, 713, 1386
Štverák, S., Maksimovic M., Travnicek P. M., Marsch E., Fazakerley A. N., & Scime E. E. 2009, JGR, 114, A05104
Suyari, H. 2006, Physica A, 368, 63
Swaczyna, P., McComas, D J., & Schwadron, N.A. 2019, ApJ, 871, 254
Tetrode, O. 1912, Ann. Phys., 38, 434
Treumann, R.A. 1997, GRL, 1997, 24, 1727
Tsallis, C. 1988, J. Stat. Phys., 52, 479
Tsallis, C. 2019, Entrp, 21, 696
Vasyliũnas, V.M., 1968, JGRA, 73, 2839
Vocks, C., Mann, G., & Rausche, G. 2008, A&A, 480, 527
Wang, C.-P., Lyons, L. R., Chen, M. W., Wolf, R. A., & Toffoletto, F. R. 2003, JGRA, 108, 1074
Xiao, F., C. Shen, Y. Wang, H. Zheng, & S. Whang 2008, JGRA, 113, A05203.
Yao, Z.-W., & Zhang, Y. 2022, ApJ, 936, 143

Yoon, P. H., Ziebell, L. F., Gaelzer, R., Lin, R. P., & Wang, L. 2012, SSRv, 173, 459
Yoon, P.H. 2014, JGRA, 119, 7074
Yoon, P.H. 2019, Classical kinetic theory of weakly turbulent nonlinear plasma processes, Cambridge Univ. Press: Cambridge
Zank, G.P., Li, G., Florinski, V., Hu, Q., Lario, D., & Smith, C.W. 2006, JGRA, 111, A06108
Zank, G. P, Heerikhuisen, J., Pogorelov, N. V., Burrows, R., McComas, D. 2010, ApJ, 708, 1092
Zank, G. P. 2015, Ann. Rev. Astron. Astrophys., 53, 449
Zhang, Y., Liu, X.-W., & Zhang B. 2014, ApJ, 780, 93
Zirnstein, E. J., & McComas, D. J. 2015, ApJ, 815, 31
Zouganelis, I. 2008, JGR, 113, A08111

## Appendix A. Proportionality constant in entropic function

Here, we calculate the proportionality constant involved in the derived entropic function in Eq.(28), i.e., $f(x) = const. \cdot \kappa \cdot (x - x^{1+\frac{1}{\kappa}})$, where we find that const. = 1. We achieve to this conclusion, through the following steps: (a) express the distribution function in terms of the entropic function $f$; (b) derive the mean kinetic energy, using the distribution function found in (a); and (c) find an equation that involves the entropy proportionality constant, and solve it.

*(a) Distribution Function in terms of f.*

Starting from this entropic function $f$, its inverse-derivative function is given by:

$$f(x) = (Ac^{-1})^{1+\frac{1}{\kappa}} \cdot \kappa \cdot (x - x^{1+\frac{1}{\kappa}})$$
$$\Leftrightarrow \tag{A1}$$
$$f'^{-1}(x - \lambda_1) = (1 + \tfrac{1}{\kappa})^{-\kappa} \cdot \left[1 - \frac{x - \lambda_1}{(Ac^{-1})^{1+\frac{1}{\kappa}} \cdot \kappa}\right]^{\kappa}.$$

This inverse-derivative function provides the energy distribution function, according to Eq.(14), i.e.,

$$p(\varepsilon) \propto f'^{-1}(k \cdot \beta\varepsilon - \lambda_1) = f'^{-1}(x - \lambda_1), \text{ with } x \equiv k\beta\varepsilon. \tag{A2}$$

*(b) Mean energy.*

Using the distribution function, shown in Eq.(A2), we derive the mean kinetic energy, $\langle \varepsilon \rangle$, or equivalently, $\langle x \equiv k\beta\varepsilon \rangle$. This is given by

$$\int_0^\infty (x - \lambda_1) f'^{-1}(x - \lambda_1) x^{\frac{1}{2}d-1} dx = (1 + \tfrac{1}{\kappa})^{-\kappa}$$
$$\times \int_0^\infty (x - \lambda_1) \left[1 - \frac{x - \lambda_1}{(Ac^{-1})^{1+\frac{1}{\kappa}}\kappa}\right]^{\kappa} x^{\frac{1}{2}d-1} dx. \tag{A3}$$

We can write

$$\int_0^\infty (x - \lambda_1) \left[1 - \frac{x - \lambda_1}{(Ac^{-1})^{1+\frac{1}{\kappa}}\kappa}\right]^{\kappa} x^{\frac{1}{2}d-1} dx = (Ac^{-1})^{1+\frac{1}{\kappa}} \kappa$$
$$\times \int_0^\infty \left\{\left[1 - \frac{x - \lambda_1}{(Ac^{-1})^{1+\frac{1}{\kappa}}\kappa}\right]^{\kappa} - \left[1 - \frac{x - \lambda_1}{(Ac^{-1})^{1+\frac{1}{\kappa}}\kappa}\right]^{\kappa+1}\right\} x^{\frac{1}{2}d-1} dx. \tag{A4}$$

Namely,

$$\int_0^\infty (x - \lambda_1) f'^{-1}(x - \lambda_1) x^{\frac{1}{2}d-1} dx = (1 + \tfrac{1}{\kappa})^{-\kappa} (Ac^{-1})^{1+\frac{1}{\kappa}} \kappa$$
$$\times \int_0^\infty \left\{\left[1 - \frac{x - \lambda_1}{(Ac^{-1})^{1+\frac{1}{\kappa}}\kappa}\right]^{\kappa} - \left[1 - \frac{x - \lambda_1}{(Ac^{-1})^{1+\frac{1}{\kappa}}\kappa}\right]^{\kappa+1}\right\} x^{\frac{1}{2}d-1} dx. \tag{A5}$$

We derive two relations to substitute into the two integrated terms in Eq.(A5). First, from Eq.(16), we have

$$\int_0^\infty \left[1 - \frac{x - \lambda_1}{(Ac^{-1})^{1+\frac{1}{\kappa}} \cdot \kappa}\right]^\kappa x^{\frac{1}{2}d-1} dx = A g_1^{-1} \cdot (1+\tfrac{1}{\kappa})^\kappa. \tag{A6}$$

while from Eq.(23), we find

$$\int_0^\infty \left[1 - \frac{x - \lambda_1}{(Ac^{-1})^{1+\frac{1}{\kappa}} \cdot \kappa}\right]^{\kappa+1} \cdot x^{\frac{1}{2}d-1} dx = c^{1+\frac{1}{\kappa}} \cdot (1+\tfrac{1}{\kappa})^{\kappa+1} \cdot g_1^{-1}. \tag{A7}$$

Therefore,

$$\int_0^\infty \left\{\left[1 - \frac{x - \lambda_1}{(Ac^{-1})^{1+\frac{1}{\kappa}} \cdot \kappa}\right]^\kappa - \left[1 - \frac{x - \lambda_1}{(Ac^{-1})^{1+\frac{1}{\kappa}} \cdot \kappa}\right]^{\kappa+1}\right\} \cdot x^{\frac{1}{2}d-1} dx$$
$$= \left[A - c^{1+\frac{1}{\kappa}}(1+\tfrac{1}{\kappa})\right] \cdot (1+\tfrac{1}{\kappa})^\kappa \cdot g_1^{-1} \tag{A8}$$

Then, Eq.(A5) becomes

$$\int_0^\infty (x - \lambda_1) \cdot f'^{-1}(x - \lambda_1) \cdot x^{\frac{1}{2}d-1} dx = \left[\kappa A \cdot (Ac^{-1})^{1+\frac{1}{\kappa}} - (\kappa+1) A^{1+\frac{1}{\kappa}}\right] \cdot g_1^{-1}. \tag{A9}$$

*(c) Solving for the entropy proportionality constant.*

Using the mean energy integral, derived in Eq.(A9), we create an equation for finding the proportionality constant.

We observe that for $y=1$, Eq.(21) is much simplified expressed, i.e.,

$$\left[\kappa A - (\kappa+1) A^{1+\frac{1}{\kappa}}\right] \cdot g_1^{-1} = \int_0^\infty (x - \lambda_1) \cdot f'^{-1}(x - \lambda_1) \cdot x^{\frac{1}{2}d-1} dx, \tag{A10}$$

where the right-hand side equals exactly the mean energy integral, derived in Eq.(A9). Hence, we obtain

$$\left[\kappa A - (\kappa+1) A^{1+\frac{1}{\kappa}}\right] \cdot g_1^{-1} = \left[\kappa A \cdot (Ac^{-1})^{1+\frac{1}{\kappa}} - (\kappa+1) A^{1+\frac{1}{\kappa}}\right] \cdot g_1^{-1}, \tag{A11}$$

leading to

$$(Ac^{-1})^{1+\frac{1}{\kappa}} = 1 \text{ or } Ac^{-1} = 1. \tag{A12}$$